\documentclass[11pt,letterpaper,]{article}
\usepackage{lmodern}
\usepackage{amssymb,amsmath}
\usepackage{ifxetex,ifluatex}
\usepackage{fixltx2e} % provides \textsubscript
\ifnum 0\ifxetex 1\fi\ifluatex 1\fi=0 % if pdftex
  \usepackage[T1]{fontenc}
  \usepackage[utf8]{inputenc}
\else % if luatex or xelatex
  \ifxetex
    \usepackage{mathspec}
  \else
    \usepackage{fontspec}
  \fi
  \defaultfontfeatures{Ligatures=TeX,Scale=MatchLowercase}
\fi
\IfFileExists{upquote.sty}{\usepackage{upquote}}{}
\IfFileExists{microtype.sty}{%
\usepackage{microtype}
\UseMicrotypeSet[protrusion]{basicmath} % disable protrusion for tt fonts
}{}
\usepackage{hyperref}
\hypersetup{
            pdftitle={A causation coefficient and taxonomy of correlation/causation relationships},
            pdfauthor={Joshua Brul\'e},
            pdfborder={0 0 0},
            breaklinks=true}
\usepackage[style=ieee,]{biblatex}
\usepackage{longtable,booktabs}
\usepackage{graphicx,grffile}
\makeatletter
\def\maxwidth{\ifdim\Gin@nat@width>\linewidth\linewidth\else\Gin@nat@width\fi}
\def\maxheight{\ifdim\Gin@nat@height>\textheight\textheight\else\Gin@nat@height\fi}
\makeatother
\setkeys{Gin}{width=\maxwidth,height=\maxheight,keepaspectratio}
\IfFileExists{parskip.sty}{%
\usepackage{parskip}
}{% else
\setlength{\parindent}{0pt}
\setlength{\parskip}{6pt plus 2pt minus 1pt}
}
\providecommand{\tightlist}{%
  \setlength{\itemsep}{0pt}\setlength{\parskip}{0pt}}
\ifx\paragraph\undefined\else
\let\oldparagraph\paragraph
\renewcommand{\paragraph}[1]{\oldparagraph{#1}\mbox{}}
\fi
\ifx\subparagraph\undefined\else
\let\oldsubparagraph\subparagraph
\renewcommand{\subparagraph}[1]{\oldsubparagraph{#1}\mbox{}}
\fi

\title{A causation coefficient and taxonomy of correlation/causation
relationships}
\author{Joshua Brul\'e\footnote{Department of Computer Science, University of
  Maryland, College Park. \texttt{jbrule@cs.umd.edu}}}
\date{}

\begin{document}
\maketitle
\begin{abstract}
This paper introduces a causation coefficient which is defined in terms
of probabilistic causal models. This coefficient is suggested as the
natural causal analogue of the Pearson correlation coefficient and
permits comparing causation and correlation to each other in a simple,
yet rigorous manner. Together, these coefficients provide a natural way
to classify the possible correlation/causation relationships that can
occur in practice and examples of each relationship are provided. In
addition, the typical relationship between correlation and causation is
analyzed to provide insight into why correlation and causation are often
conflated. Finally, example calculations of the causation coefficient
are shown on a real data set.
\end{abstract}

\section{Introduction}\label{introduction}

The maxim, ``Correlation is not causation'', is an important warning to
analysts, but provides very little information about what causation is
and how it relates to correlation. This has prompted other attempts at
summarizing the relationship. For example, Tufte \autocite{tufte2006}
suggests either, ``Observed covariation is necessary but not sufficient
for causality'', which is demonstrably false or, ``Correlation is not
causation but it is a hint'', which is correct, but still
underspecified. In what sense is correlation a `hint' to causation?

Correlation is well understood and precisely defined. Generally
speaking, correlation is any statistical relationship involving
dependence, i.e.~the random variables are not independent. More
specifically, correlation can refer to a descriptive statistic that
summarizes the nature of the dependence. Such statistics do not provide
all of the information available in the joint probability distribution,
but can provide a valuable summary that is easier to reason about. Among
the most popular, and often referred to as just ``the correlation
coefficient'' \autocite{weisstein-correlation}, is the Pearson
correlation coefficient, which is a measure of the linear correlation
between variables.

Causality is an intuitive idea that is difficult to make precise. The
key contribution of this paper is the introduction of a ``causation
coefficient'', which is suggested as the natural causal analogue of the
Pearson correlation coefficient. The causation coefficient permits
comparing correlation and causation to each other in a manner that is
both rigorous and consistent with common intuition.

The rest of this paper is outlined as follows: The statistical/causal
distinction is discussed to provide background. The existing
probabilistic causal model approach to causality is briefly summarized.
The causation coefficient is defined in terms of probabilistic causal
models and some of the properties of the coefficient are discussed to
support the claim that it is the natural causal analogue of the Pearson
correlation coefficient.

The definition of the causation coefficient permits the following new
analyses to be conducted: A taxonomy of the possible relationships
between correlation and causation is introduced, with example models.
The typical relationship between correlation and causation is analyzed
to provide insight into why correlation and causation are often
conflated. Finally, example calculations of the correlation coefficient
are shown on a real data set.

\section{Statistical/causal
distinction}\label{statisticalcausal-distinction}

Causality is difficult to formalize. Causality is implicit in the
structure of ordinary language \autocite{brown1983} and the words
`causality' and `causal' are often used to refer to a number of
disparate concepts. In particular, much confusion stems from conflating
three distinct tasks in causal inference \autocite{heckman2005}:

\begin{enumerate}
\def\labelenumi{\arabic{enumi}.}
\tightlist
\item
  Definitions of counterfactuals
\item
  Identification of causal models from population distributions
\item
  Selection of causal models given real data
\end{enumerate}

Counterfactuals, as defined in philosophy, are hypothetical or potential
outcomes -- statements about possible alternatives to the actual
situation \autocite{lewis1973}. A classic example is, ``If Nixon had
pressed the button, there would have been a nuclear holocaust''
\autocite{fine1975}, a statement which seems intuitively correct, but
difficult to formally model and impossible to empirically verify.
Defining causation in terms of counterfactuals originates with Hume in
defining a cause to be, ``where, if the first object had not been, the
second never had existed'' \autocite{hume1748}. Indeed, in a world where
there had been a nuclear war during the Nixon administration, it would
be quite reasonable to claim that launching nuclear missiles was a
cause. A key difficulty in making this notion of causality precise is
that it requires precise models of counterfactuals and therefore precise
assumptions that can be unobservable and untestable even in principle.

For example, consider the possible results of treating a patient in a
clinical setting. In the notation of the Rubin causal model
\autocite{holland1986}, a particular patient or \emph{unit}\footnote{``Units''
  are the basic objects (primitives) of study in an investigation in the
  Rubin causal model approach. Examples of units include individual
  human subjects, households, or plots of land.}, \(u\), can be
potentially exposed to either treatment, \(t\), or control, \(c\). The
\emph{treatment effect}\footnote{This is also referred to as ``causal
  effect'' in the literature. In this paper, ``treatment effect'' is
  used to avoid confusion with the related but distinct definition of
  causal effect in the probabilistic causal model approach.},
\(Y_t(u) - Y_c(u)\), is the difference between the outcomes when the
patient is exposed to treatment and when the same patient is exposed to
the control. Determining treatment effect on a unit is usually the
ultimate goal of causal inference. It is also impossible to observe --
the same patient cannot be treated and not treated -- a problem which
Holland names the \emph{Fundamental Problem of Causal Inference}.

This is not to suggest that causal inference is impossible, merely that
additional assumptions must be made if causal conclusions are to be
reached. A well known assumption that makes causal inference possible is
randomization. Assuming that units are randomly assigned to treatment or
control groups, it is possible to estimate the \emph{average treatment
effect}, \(E[Y_t - Y_c]\). Note that randomization is an assumption
external to the data; it is not possible to determine, from the data
alone, that it was obtained from a randomized controlled trial. Another
example of an assumption that permits causal inference is \emph{unit
homogeneity}, which can be thought of as ``laboratory conditions''. If
different units are carefully prepared, it may be reasonable to assume
that they are equivalent in all relevant aspects, i.e.
\(Y_t(u_1) = Y_t(u_2)\) and \(Y_c(u_1) = Y_c(u_2)\). For example, it is
often assumed that any two samples of a given chemical element are
effectively identical. In these cases, treatment effect can be
calculated directly as \(Y_t(u_1) - Y_c(u_2)\).

A closely related concept is \emph{ceteris paribus}, roughly, ``other
things held constant'', which is a mainstay of economic analysis
\autocite{heckman2005}. For example, increasing the price of some good
will cause demand to fall, assuming that no other relevant factors
change at the same time. This is not to suggest that no other factors
will change in a real economy; \emph{ceteris paribus} simply isolates
the effect of one particular change to make it more amenable to
analysis.

In practice, the first causal inference task, defining counterfactuals,
requires having a scientific theory. For example, classical mechanics
describes the possible states of idealized physical systems and can
provide an account of manipulation. The theory can predict what would
happen to the trajectory of some object if an external force were to be
applied, whether or not such a force was actually applied in the real
world. Scientific theories are usually parameterized; one example of a
parameter is standard gravity, \(g_n \approx 9.8\) \(m/s^2\), the
acceleration of an object due to gravity near the surface of the earth
\autocite{nist2008}.

The second causal inference task, identification from population
distributions, is a problem of uniquely determining a causal model or
some property of a causal model from hypothetical population data. In
other words, the problem is to find unique mappings from population
distributions or other population measures to causal parameters. This
can be thought of as the problem of determining which scientific theory
is correct, given data without sampling error. A well-designed
experiment to determine \(g_0\) will, in the limit of infinite samples,
yield the exact value for the parameter.

The third task, selection of causal models given real data, is the
problem of inference in practice. Any real experiment can only provide
an analyst with a finite-sample distribution subject to random error.
This problem lies in the domain of estimation theory and hypothesis
testing.

In addition to the standard population/sample distinction, this paper
follows Pearl's conventions in referring to the statistical/causal
distinction\footnote{This distinction has been referred to by many
  different names, including: descriptive/etiological,
  descriptive/generative, associational/causal, empirical/theoretical,
  observational/experimental, observational/interventional.}
\autocite{pearl2009}. A \emph{statistical} concept is a concept that is
definable in terms of a joint probability distribution of observed
variables. Variance is an example of a statistical parameter; the
statement that \(f\) is multivariate normal is an example of a
statistical assumption. A \emph{causal} concept is a nonstatistical
concept that is definable in terms of a causal model. Randomization is
an example of a causal, not statistical, assumption because it is
impossible to determine from a joint probability distribution that a
variable was randomly assigned.

This distinction draws a sharp line between statistical and causal
analysis, which can be thought of as the difference between analyzing
uncertain, yet static conditions versus changing conditions
\autocite{pearl2001}. Estimating and updating the likelihood of events
based on observed evidence are statistical tasks, given that
experimental conditions remain the same. Causal analysis aims to infer
the likelihood of events under changing conditions, brought about by
external interventions such as treatments or policy changes. Much like
how statistical inference is performed with respect to assumptions
formalized in a statistical model, rigorous causal inference requires
formal causal models.

\section{Probabilistic causal models}\label{probabilistic-causal-models}

Probabilistic causal models\footnote{Also referred to as ``graphical
  causal models'' and ``structural causal models'' in the literature.}
are an approach to causality characterized by nonparametric models
associated with a type of directed acyclic graph (DAG) called a causal
diagram. The concept of using graphs to model probabilistic and causal
relationships originates with Wright's path analysis
\autocite{wright1934}. The modern, nonparametric version appears to have
been first proposed by Pearl and Verma \autocite{pearl1991} and has been
the subject of considerable research since then \autocite{pearl1995}
\autocite{shpitser2008} \autocite{bareinboim2014-transportability}. This
section is meant to provide a high-level summary of probabilistic causal
models, sufficient to explain the proposed causation coefficient.

\subsection{Causal models}\label{causal-models}

The philosophy of probabilistic causal models is that of Laplacian
quasi-determinism -- a complete description of the state of a system is
sufficient to exactly determine how the system will evolve
\autocite{laplace1814}. In this view, randomness is a statement of an
analyst's ignorance, not inherent to the system itself.\footnote{This
  excludes quantum-mechanical systems from analysis. Arguably, the
  `intrinsic randomness' of these systems are why they are so often
  considered counterintuitive.}

A causal model, \(M\), consists of a set of equations where each
child-parent family is represented by a deterministic function:

\[
X_i = f_i(pa_i, \epsilon_i) \quad (i=1,\ldots,n)
\]

The \(X_i\) variables are the \emph{endogenous} variables, determined by
factors in the model. \(pa_i\) denotes the \emph{parents} of \(X_i\),
which can be thought of as the direct, known causes of \(X_i\). The
\(\epsilon_i\) variables are the \emph{exogenous} variables, and can
appropriately be considered `background' variables or `error terms' and
correspond to variables that are determined by factors outside of the
model \autocite{pearl2009}. Since the exogenous variables model those
factors that cannot be directly accounted for, they are treated as
random variables. Regardless of the distribution of the exogenous
variables, or the functional form of the \(f_i\) equations, a
probability distribution, \(P(\mathbf{\epsilon})\) over the exogenous
variables induces a probability distribution, \(P(x_1, \ldots, x_n)\)
over the endogenous variables \autocite{pearl2009}. The resulting model
is called a \emph{probabilistic causal model}.

Each causal model induces a causal diagram, \(G\), where each \(X_i\)
corresponds to a vertex and each parent-child relationship between
\(pa_i\) and \(X_i\) corresponds to a directed edge from parent to
child. In this paper, it is assumed that all models are
\emph{recursive}, i.e.~all models induce an acyclic causal diagram.

This paper adopts the convention that all of the exogenous variables are
mutually independent. If all of the endogenous variables are observable
-- denoted in the causal diagram by solid nodes -- then the model is
called \emph{Markovian}. The joint probability function,
\(P(x_1, \ldots, x_n)\), in a Markovian model is said to be \emph{Markov
compatible} with \(G\) in that \(P(x_1, \ldots, x_n)\) respects the
\emph{Markov condition}: each variable is independent of all its
non-descendants given its parents in the graph
\autocite{bareinboim2012-local}. Dependence between two observable
variables that have no observable ancestor can be introduced by adding a
\emph{latent} endogenous variable, denoted in a causal diagram by an
open node. Such a model is called \emph{semi-Markovian}.

Each causal diagram can be thought of as denoting a set of causal
models. Most of this paper considers the following set of models with
endogenous variables \(X\), \(Y\) and \(Z\):

\[Z = f_Z(\epsilon_Z)\] \[X = f_X(Z, \epsilon_X)\]
\[Y = f_Y(X, Z, \epsilon_Y)\]

\begin{figure}[htbp]
\centering
\includegraphics{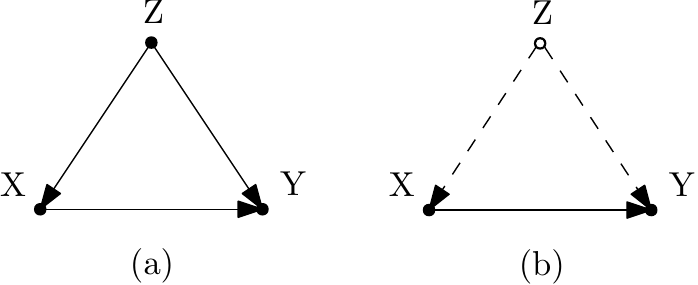}
\caption{Causal diagrams for Markovian and semi-Markovian models}
\end{figure}

\subsection{Causal effect}\label{causal-effect}

Without any additional context, this characterization of probabilistic
causal models appears merely to be a way to generate Bayesian networks.
However, the functional, quasi-deterministic approach also specifies how
the probability distribution of the observable variables change in
response to an external intervention. The simplest external intervention
is where a single variable \(X_i\) is forced to take some fixed value
\(x_i\), `setting' or `holding constant' \(X_i = x_i\). Such an
\emph{atomic} intervention corresponds to replacing the equation
\(X_i = f_i(pa_i, \epsilon_i)\) with the constant \(x_i\), generating a
new model. This can be extended to sets of variables.

\begin{description}
\tightlist
\item[Causal effect]
\autocite{pearl1995} Given two disjoint sets of variables, \(X\) and
\(Y\), the \emph{causal effect} of \(X\) on \(Y\), denoted either as
\(P(y \mid \hat{x})\) or \(P(y \mid do(x))\) is a function from \(X\) to
the space of probability distributions on \(Y\). For each realization
\(x\) of \(X\), \(P(y \mid \hat{x})\) gives the probability of \(Y=y\)
induced by deleting from the model all equations corresponding to
variables in \(X\) and substituting \(X=x\) in the remaining equations.
\end{description}

Crucially, causal effect, \(P(y \mid \hat{x})\), is fundamentally
different than conditioning or observation, \(P(y \mid x)\). The latter
is a function of the joint probability distribution of the original
model, \(M\). The former is a function of the distribution of the
submodel, \(M_x\), that results from the effect of action \(do(X=x)\) on
\(M\). Intuitively, this can be thought of as `cutting' all of the
incoming edges to \(X\) and replacing the random variable with the
constant \(x\).

\begin{figure}[htbp]
\centering
\includegraphics{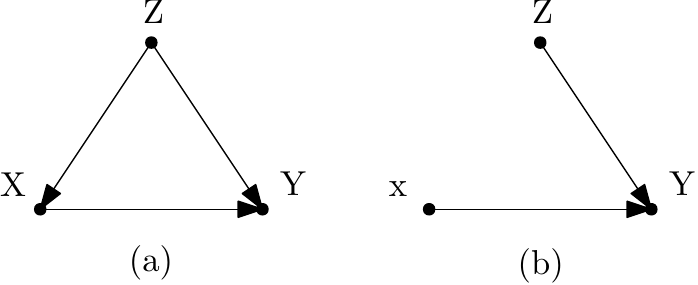}
\caption{An intervention \(do(X=x)\) on causal model \(M\) produces
submodel \(M_x\)}
\end{figure}

It is possible for DAGs to be \emph{observationally equivalent},
i.e.~Markov compatible with the same set of joint probability
distributions \autocite{pearl1991}. Two observationally equivalent DAGs
cannot be distinguished without performing interventions or drawing on
additional causal information. For example \autocite{pearl2009}, in a
causal diagram modeling relationships between the season, rain,
sprinkler settings and whether the ground is wet, it would be reasonable
to accept a model where the season causally effects the sprinkler
settings, but not vise-versa. While indistinguishable from observation
alone, the two models imply different results due to intervention; it
would be implausible that changing the settings on a sprinkler would
cause the season of the year to change as a result.

\begin{figure}[htbp]
\centering
\includegraphics{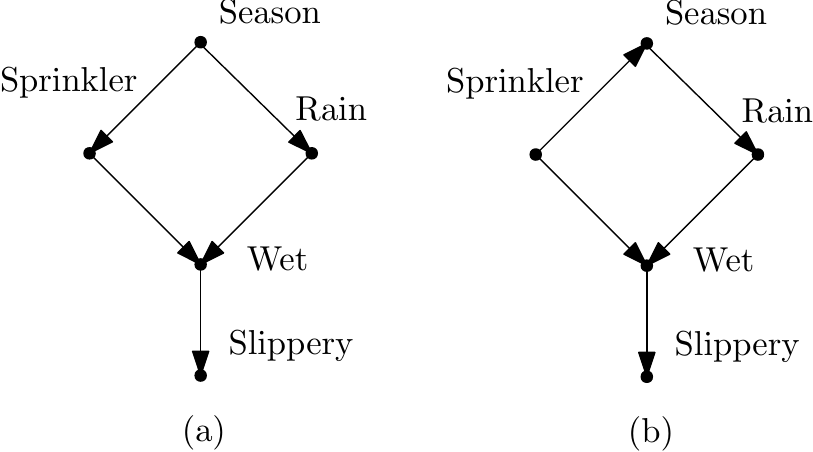}
\caption{Observationally equivalent DAGs}
\end{figure}

\subsection{Identification of causal
effect}\label{identification-of-causal-effect}

The problem of whether a causal query can be uniquely answered is
referred to as causal \emph{identifiability}. An unbiased estimate of
\(P(y \mid \hat{x}_i)\) can always be calculated from observational
(preintervention) probabilities in Markovian models by conditioning on
the parents of \(X_i\) and averaging the result, weighted by the
probabilities of \(PA_i = pa_i\). This operation is called ``adjusting
for \(PA_i\)'' or ``adjustment for direct causes'' \autocite{pearl2009}.
More formally, the observational probability distribution \(P\) and
causal diagram \(G\) of a Markovian model \emph{identifies} the effect
of the intervention \(do(X_i = x_i)\) on \(Y\) and is given by:

\[P(y \mid \hat{x}_i) = \sum_{pa_i} P(y \mid x_i, pa_i) P(pa_i)\]

Semi-Markovian models do not always permit identification. A simple
example is when a single latent variable is a parent of every observable
variable. Informally, it is not possible to determine if observed
covariation is indicative of a causal effect between two variables, or
whether their common, unobservable parent brings about the correlation.
For example \autocite{shpitser2008}, consider two models, \(M^1\) and
\(M^2\) where both models have observable variables \(X, Y\), latent
\(Z\sim\text{Bernoulli}(0.5)\), and \(f_X(z) = z\). In \(M^1\),
\(f_Y(z, x) = z \text{ XOR } x\); in \(M^2\), \(f_Y(z, x) = 0\). These
models are compatible with the same causal diagrams and have identical
observational probability distributions, but different causal effects,
\(P(y \mid \hat{x})\). Since the causal effect cannot be uniquely
calculated from the available information, it is not identifiable.
However, many semi-Markovian models still permit estimation of certain
causal effects. Complete methods are described in
\autocite{shpitser2008}.

\section{The causation coefficient}\label{the-causation-coefficient}

The Pearson product-moment correlation coefficient, \(\rho\), is a
standard measure of correlation between random variables. This is
commonly described as a measure of how well the relationship between
\(X\) and \(Y\) can modeled by a linear relationship with \(\rho=-1/+1\)
being a perfect negative/positive linear relationship and \(0\)
representing no linear relationship at all. The population correlation
coefficient is defined as a normalized covariance
\autocite{weisstein-correlation}:

\[
\rho_{X,Y} = \frac{cov(X, Y)}{\sqrt{Var[X] Var[Y]}} = \frac{E[XY] - E[X] E[Y]}{\sqrt{(E[X^2] - E[X]^2) (E[Y^2] - E[Y]^2)}}
\]

For discrete random variables, this is a function of the joint
probability mass function (for continuous random variables that admit a
probability density function, the summations are replaced with
integrals):

\[
\rho_{X,Y} = \frac{\sum_x \sum_y xy P(x,y) - \sum_x x P(x) \sum_y y P(y)}{\sqrt{(\sum_x x^2 P(x) - (\sum_x x P(x))^2) (\sum_y y^2 P(y) - (\sum_y y P(y))^2)}}
\]

The causation coefficient relies on the observation that the correlation
coefficient can, by the law of total probability, be rewritten as a
function of the conditional distribution \(P(y \mid x)\), and marginal
distribution, \(P(x)\), instead of in terms of the joint density:

\[
\rho_{X,Y} = \frac{\sum_x \sum_y xy P(y \mid x) P(x) - \sum_x x P(x) \sum_x \sum_y y P(y \mid x) P(x)}{\sqrt{Var[X] (\sum_x \sum_y y^2 P(y \mid x) P(x) - (\sum_x \sum_y y P(y \mid x) P(x))^2)}}
\]

Syntactically, the causation coefficient, \(\gamma_{X \to Y}\), is
defined by replacing \(P(y | x)\) with \(P(y | \hat{x})\) and \(P(x)\)
with \(\hat{P}(x)\). As a convenience, the following terms are also
defined:
\(Var[\hat{X}] = \sum_x x^2 \hat{P}(x) - (\sum_x x \hat{P}(x))^2\) and
\(Var[Y_{\hat{X}}] = \sum_x \sum_y y^2 P(y | \hat{x}) \hat{P}(x) - (\sum_x \sum_y y P(y | \hat{x}) \hat{P}(x))^2\).
The full definition of \(\gamma_{X \to Y}\) is then:

\[
\gamma_{X \to Y} = \frac{\sum_x \sum_y xy P(y \mid \hat{x}) \hat{P}(x) - \sum_x x \hat{P}(x) \sum_x \sum_y y P(y \mid \hat{x}) \hat{P}(x)}{\sqrt{Var[\hat{X}] Var[Y_{\hat{X}}]}}
\]

Where \(P(y \mid \hat{x})\) is the causal effect of \(do(X=x)\) on \(Y\)
and \(\hat{P}(x)\) is the \emph{distribution of interventions}.

\subsection{Distribution of
interventions}\label{distribution-of-interventions}

In the discrete case, the distribution of interventions can be thought
of as a set of weights for averaging the possible causal effects. It
also has an interpretation in the context of observational and
experimental studies. As an example, consider a scenario where patients
decide for themselves whether or not to take a drug (\(X\)), and observe
whether or not they recover (\(Y\)). The population joint probability
distribution, \(P(x, y)\), provides all of the information available
from an idealized version of this observational study. For intuition, it
may be helpful to imagine \(P(x, y)\) as being calculated from millions
of samples to the point where random sampling error has ceased to be a
relevant consideration.

This simplest way to model this is with Bernoulli (binary) random
variables for \(X\) and \(Y\), with \(0\) representing no treatment or
failure to recover and \(1\) representing treatment or recovery. The
probability of patients deciding for themselves whether or not to take
the drug, in this observational study, is the marginal probability
\(P(x)\). In clinical terms, \(P(X=0)\) and \(P(X=1)\) are the relative
sizes of the cohorts.

However, even in an idealized observational study, \(P(y \mid x)\) would
not provide definitive information on whether treatment actually
improves patient outcomes. Hypothetically, the drug could cause
unpleasant side effects in the patients that would have received the
greatest benefit, leading those patients to choose not to take the drug.
An idealized randomized controlled trial would permit an analyst to
directly measure \(P(y \mid \hat{x})\), as randomization explicitly cuts
out confounding. However, randomized controlled trials are often
impractical (e.g.~too expensive or unethical) to run in practice.

The relative sizes of the cohorts in an observational study may be
different than the relative sizes of the treatment and control groups in
a corresponding randomized controlled trial -- this is the use of the
distribution of interventions \(\hat{P}\). Experiments are often
designed to have equal group sizes as this typically provides maximum
statistical power, but this is by no means universal. Also, it is not
uncommon for patients to drop out or otherwise be disqualified from
studies, so the cohorts will often be unequal in practice.

The \emph{natural causation coefficient}, denoted \(\gamma_{X \to Y}\)
or \(\gamma\), is defined for \(\hat{P}(x)\) equal to the
pre-intervention marginal distribution, \(P(x)\). This corresponds to an
experimental trial where the treatment groups are scaled to be
proportional to the relative sizes seen in the observational study.

The \emph{maximum entropy causation coefficient}, denoted
\(\gamma_{H, X \to Y}\) or \(\gamma_{H}\), is the causation coefficient
where \(\hat{P}(x)\) is a maximum entropy probability distribution. For
random variables with bounded support, this is the uniform distribution
and corresponds to equal treatment group sizes.

Other distributions of interventions are possible, to reweigh the
effects of certain interventions relative to others in the computation
of the causation coefficient. These should be denoted explicitly as
\(\gamma_{\hat{P}}\). For example, a certain drug may be known to be
helpful in certain small doses, but worse than no treatment at all in
larger doses, in which case both the natural and maximum entropy
coefficients could be misleading. In such cases, a distribution of
interventions corresponding to current best practices may be more
informative.

\subsection{Independence and
invariance}\label{independence-and-invariance}

The definition of \emph{independence} of random variables \(X\) and
\(Y\) is: \(\forall x, y \enskip P(x, y) = P(x) P(y)\) or, equivalently:
\(\forall x,y \enskip P(y \mid x)=P(y)\). In other words, observing
\(X\) provides no information about \(Y\) (and vise-versa). The causal
equivalent is \emph{invariance} of \(Y\) to \(X\):
\(\forall x, y \enskip P(y \mid \hat{x}) = P(y)\); that is to say, no
possible intervention on \(X\) can affect \(Y\)
\autocite{bareinboim2012-local}. Unlike independence, invariance is not
symmetric. The term \emph{mutually invariant} is suggested to refer to
when both \(Y\) is invariant to \(X\) and \(X\) is invariant to \(Y\).

For Bernoulli random variables, \(X\) and \(Y\) are uncorrelated
(\(\rho=0\)) if and only if they are independent. The analogous
condition holds for the causation coefficient. For Bernoulli distributed
\(X\) and \(Y\), \(\gamma_{X \to Y}=0\) if and only if \(Y\) is
invariant to \(X\) (see appendix for proof). However, both the
correlation and causation coefficients have difficulty capturing
nonlinear relationships between variables.\footnote{Also worth noting is
  that neither coefficient is robust to outliers. This can be mitigated
  by winsorizing or trimming.} In general, independence implies
\(\rho=0\) and invariance implies \(\gamma=0\), but the converse does
not hold for many distributions.

\begin{longtable}[]{@{}lll@{}}
\caption{Non-invariant interventional distributions where \(\gamma_H=0\)
\label{non-invariant}}\tabularnewline
\toprule
\(P(y \mid \hat{x})\) & y=0 & y=1\tabularnewline
\midrule
\endfirsthead
\toprule
\(P(y \mid \hat{x})\) & y=0 & y=1\tabularnewline
\midrule
\endhead
x=-1 & 1/3 & 2/3\tabularnewline
x=0 & 2/3 & 1/3\tabularnewline
x=1 & 1/3 & 2/3\tabularnewline
\bottomrule
\end{longtable}

As a simple example, Table \ref{non-invariant} contains interventional
distributions where \(Y\) is not invariant to \(X\), but the maximum
entropy causation coefficient \(\gamma_H=0\). The natural causation
coefficient may be positive, negative or zero depending on the
observational (pre-intervention) distribution \(P(x)\).

\subsection{Average treatment effect}\label{average-treatment-effect}

Average treatment effect is defined as \autocite{chickering1996}:

\[
ATE(X \to Y) = P(Y = 1 \mid do(X = 1)) - P(Y = 1 \mid do(X=0))
\]

This is the probabilistic causal model equivalent of the Rubin causal
model definition of average treatment effect. Positive ATE implies that
treatment is, on average, superior to non-treatment, while negative ATE
implies the opposite. For Bernoulli distributed \(X\) and \(Y\),
\(\gamma_{X \to Y}\) reduces to (see appendix for proof):

\[
\gamma_{X \to Y} = \text{ATE}(X \to Y) \sqrt{\frac{Var[\hat{X}]}{Var_{\hat{X}}[Y]}}
\]

Since variance is strictly positive for nondegenerate Bernoulli
distributions, this implies that \(\gamma\) has the same sign as the
average treatment effect.

\section{A taxonomy of correlation/causation
relationships}\label{a-taxonomy-of-correlationcausation-relationships}

For Bernoulli \(X\) and \(Y\), \(\rho\) and \(\gamma\) provide a natural
way to classify the possible correlation/causation relationships.
\(\rho\) and \(\gamma\) can each be positive, negative or zero, implying
9 possible relationships. These are grouped into 5 classifications in
Table \ref{taxonomy}.

\begin{longtable}[]{@{}lll@{}}
\caption{Correlation/causation relationships
\label{taxonomy}}\tabularnewline
\toprule
& \(\rho\) & \(\gamma\)\tabularnewline
\midrule
\endfirsthead
\toprule
& \(\rho\) & \(\gamma\)\tabularnewline
\midrule
\endhead
independent and invariant & 0 & 0\tabularnewline
common causation & +/- & 0\tabularnewline
inverse causation & +/- & -/+\tabularnewline
unfaithful & 0 & +/-\tabularnewline
genuine causation & +/- & +/-\tabularnewline
\bottomrule
\end{longtable}

In Table \ref{taxonomy}, ``0'' is a zero value for the coefficient, and
``+/-'' refers to the coefficient taking on a positive or a negative
value (e.g.~inverse causation refers to either a model with positive
\(\rho\) and negative \(\gamma\), or negative \(\rho\) and positive
\(\gamma\)). Note that \(\rho\) and \(\gamma\) are population
coefficients; this taxonomy can be thought of as categorizing the
possible relationships between correlation and causation, in the limit
of infinite samples.

Many of the relationships described in the following sections are well
known and existing terminology is used where appropriate. Examples of
each relationship are given, as well as simple probabilistic causal
models of three Bernoulli distributed variables that produce the
described relationship. Notably absent is the notion of mutual
causation, which is beyond the scope of this paper. Note that while
\(\rho\) is symmetric, i.e. \(\rho_{X,Y} = \rho_{Y,X}\), at least one of
\(\gamma_{X \to Y}\), \(\gamma_{Y \to X}\) is zero in all recursive
probabilistic causal models (see appendix for proof).

\subsection{Independent and invariant}\label{independent-and-invariant}

Two variables that are independent and mutually invariant are completely
unrelated -- neither observing nor manipulating one can provide
information about or change the other. This is usually the default
assumption when studying a system -- in hypothesis testing, the null
hypothesis is usually ``no effect''. For a somewhat absurd example,
researchers would not believe that the average gas mileage of a Prius is
related in any way to the minimum width of the English channel
\autocite{munroe2010} by default -- some sort of evidence would be
expected before taking such a suggestion seriously. The notion of light
cones provides an example familiar to physicists -- the principle of
locality and the theory special relativity imply that nothing outside of
someone's past and future light cones can ever affect them.

Independent and invariant variables can be trivially mathematically
modeled. An example is provided here to introduce the conventions used
throughout the rest of this section. Let
\(\epsilon_X, \epsilon_Y, \epsilon_Z\) be fair coins, i.e.~independent
Bernoulli distributed random variables with \(p=0.5\). These are the
exogenous variables of the probabilistic causal model. \(X\) will
generally model a cause or treatment, \(Y\), an effect or response, and
\(Z\), a confounding variable that causally effects \(X\) and \(Y\). An
example model with independent and invariant \(X\) and \(Y\) is simply:

\[Z = \epsilon_Z\] \[X = \epsilon_X\] \[Y = \epsilon_Y\]

\begin{longtable}[]{@{}llll@{}}
\caption{Observational distribution of independent and invariant
model}\tabularnewline
\toprule
\(P(x,y)\) & y=0 & y=1 & \(P(x)\)\tabularnewline
\midrule
\endfirsthead
\toprule
\(P(x,y)\) & y=0 & y=1 & \(P(x)\)\tabularnewline
\midrule
\endhead
x=0 & 1/4 & 1/4 & 1/2\tabularnewline
x=1 & 1/4 & 1/4 & 1/2\tabularnewline
\(P(y)\) & 1/2 & 1/2 &\tabularnewline
\bottomrule
\end{longtable}

\begin{longtable}[]{@{}lll@{}}
\caption{Interventional distributions of independent and invariant
model}\tabularnewline
\toprule
\(P(y \mid \hat{x})\) & y=0 & y=1\tabularnewline
\midrule
\endfirsthead
\toprule
\(P(y \mid \hat{x})\) & y=0 & y=1\tabularnewline
\midrule
\endhead
x=0 & 1/2 & 1/2\tabularnewline
x=1 & 1/2 & 1/2\tabularnewline
\bottomrule
\end{longtable}

\(X\) and \(Y\) are clearly independent and invariant and the
correlation and causal coefficients are \(0\).

\subsection{Common causation}\label{common-causation}

Reichenbach appears to be the first to propose the ``Principle of the
Common Cause'' claiming, ``If an improbable coincidence has occurred,
there must exist a common cause'' \autocite{reichenbach1956}.
Elaborating on this, he suggests that correlation between events \(A\)
and \(B\) indicates either that \(A\) causes \(B\), \(B\) causes \(A\)
or \(A\) and \(B\) have a common cause. This philosophical claim
naturally suggests the following definition:

\begin{description}
\tightlist
\item[Common Causation]
\(X\) and \(Y\) are said to experience common causation when \(X\) and
\(Y\) are mutually invariant but not independent.
\end{description}

This effect is sometimes referred to as a ``spurious relationship'' or
``spurious correlation'' -- a term originally coined by Pearson
\autocite{pearson1896}. This risks conflating several distinct concepts:
the interventional distributions from which \(\gamma\) is calculated,
the population observational distribution from which \(\rho\) is
calculated, and the finite-sample observational distribution, from which
the sample correlation coefficient, \(r\) is calculated. Consider the
following scenarios:

\begin{itemize}
\item
  A very large number of samples are taken from invariant \(X\) and
  \(Y\), but due to a latent confounding variable, \(X\) and \(Y\) are
  correlated.
\item
  A small number of samples are taken from independent and invariant
  \(X\) and \(Y\), but due to random sampling errors, the sample
  correlation coefficient suggests that \(X\) and \(Y\) are correlated.
\end{itemize}

In both scenarios, there is a spurious relationship between \(X\) and
\(Y\). The first scenario exhibits common causation. The second scenario
is due to random sampling error and, as the number of samples increases,
the observed correlation will tend to zero. The term ``coincidental
correlation'' is suggested to distinguish this finite-sample effect from
common causation.

An example of a common cause can be found in a study on myopia and
ambient lighting at night \autocite{quinn1999}. Development of myopia
(shortsightedness) is correlated with nighttime light exposure in
children, although the latter does not cause the former. The common
cause is that myopic parents are likely to have myopic children, and
also more likely to set up night lights.

The following is a simple common causation model: Let
\(\epsilon_X, \epsilon_Y, \epsilon_Z\) be fair coins and \(X\), \(Y\)
and \(Z\) be defined by the following three equations:

\[Z = \epsilon_Z\] \[X = Z \land \epsilon_X\] \[Y = Z \land \epsilon_Y\]

\begin{longtable}[]{@{}llll@{}}
\caption{Observational distribution of common cause
model}\tabularnewline
\toprule
\(P(x,y)\) & y=0 & y=1 & \(P(x)\)\tabularnewline
\midrule
\endfirsthead
\toprule
\(P(x,y)\) & y=0 & y=1 & \(P(x)\)\tabularnewline
\midrule
\endhead
x=0 & 5/8 & 1/8 & 3/4\tabularnewline
x=1 & 1/8 & 1/8 & 1/4\tabularnewline
\(P(y)\) & 3/4 & 1/4 &\tabularnewline
\bottomrule
\end{longtable}

\begin{longtable}[]{@{}lll@{}}
\caption{Interventional distributions of common cause
model}\tabularnewline
\toprule
\(P(y \mid \hat{x})\) & y=0 & y=1\tabularnewline
\midrule
\endfirsthead
\toprule
\(P(y \mid \hat{x})\) & y=0 & y=1\tabularnewline
\midrule
\endhead
x=0 & 3/4 & 1/4\tabularnewline
x=1 & 3/4 & 1/4\tabularnewline
\bottomrule
\end{longtable}

From the observational distribution, it is clear that \(X\) and \(Y\)
are correlated (\(\rho=1/3\)) and from the interventional distributions,
that \(X\) and \(Y\) are invariant (\(\gamma=0\)).

\subsection{Inverse causation}\label{inverse-causation}

A classic veridical paradox is the relationship between tuberculosis and
dry climate \autocite{gardner2006}. At one point, Arizona, with one of
the driest climates in the United States was found to also have the
largest share of tuberculosis deaths. This is because tuberculosis
patients greatly benefit from a dry climate, and many moved there. The
following is proposed as a definition for this type of scenario:

\begin{description}
\tightlist
\item[Inverse causation]
\(X\) and \(Y\) are said to experience inverse causation when the
correlation coefficient \(\rho\) and causation coefficient \(\gamma\)
have the opposite sign.
\end{description}

Inverse causation is of special importance when considering clinical
treatment; a case of inverse causation is a case where the correct
treatment option is the opposite of what a naive interpretation of
correlation would suggest.

The following is a simple model that exhibits inverse causation: Let
\(\epsilon_Z\) be a fair coin, and \(\epsilon_Y\) be Bernoulli
distributed with \(p=3/4\). The following is an inverse causation model
with \(\rho=-1/2\) and \(\gamma=1/4\):

\[Z = \epsilon_Z\] \[X = Z\] \[
Y = 
\begin{cases}
\lnot Z & \text{if } \epsilon_Y=1 \\
X & \text{if } \epsilon_Y=0 
\end{cases}
\]

\begin{longtable}[]{@{}llll@{}}
\caption{Observational distribution of inverse causation
model}\tabularnewline
\toprule
\(P(x,y)\) & y=0 & y=1 & \(P(x)\)\tabularnewline
\midrule
\endfirsthead
\toprule
\(P(x,y)\) & y=0 & y=1 & \(P(x)\)\tabularnewline
\midrule
\endhead
x=0 & 1/8 & 3/8 & 1/2\tabularnewline
x=1 & 3/8 & 1/8 & 1/2\tabularnewline
\(P(y)\) & 1/2 & 1/2 &\tabularnewline
\bottomrule
\end{longtable}

\begin{longtable}[]{@{}lll@{}}
\caption{Interventional distributions of inverse causation
model}\tabularnewline
\toprule
\(P(y \mid \hat{x})\) & y=0 & y=1\tabularnewline
\midrule
\endfirsthead
\toprule
\(P(y \mid \hat{x})\) & y=0 & y=1\tabularnewline
\midrule
\endhead
x=0 & 5/8 & 3/8\tabularnewline
x=1 & 3/8 & 5/8\tabularnewline
\bottomrule
\end{longtable}

``Inverse causation'' suggested to avoid confusion with other
terminology. ``Anti-causation'' is inappropriate, as ``anti-causal
filters'' in digital signal processing are filters whose output depend
on future inputs. ``Reverse causation'' is also inappropriate, as this
refers to mistakenly believing that \(Y\) has a causal effect on \(X\),
when, \(X\) causes \(Y\).

\subsection{Unfaithfulness}\label{unfaithfulness}

The Markov condition entails a set of conditional independence relations
between variables corresponding to nodes in a DAG. The
\emph{faithfulness} condition {[}spirtes2000{]} (also referred to as
\emph{stability} \autocite{pearl2009}) is the converse.

\begin{description}
\tightlist
\item[Faithfulness condition]
A distribution \(P\) is faithful to a DAG \(G\) if no conditional
independence relations other than the ones entailed by the Markov
condition are present.
\end{description}

This is a global condition, applying to a joint probability distribution
and a DAG. The following local condition is defined in terms of two
random variables \(X\) and \(Y\) in a causal model:

\begin{description}
\tightlist
\item[Unfaithful]
\(X\) and \(Y\) are said to be unfaithful if they are independent but
not mutually invariant.
\end{description}

This local condition can only occur if the global faithfulness condition
is violated (see appendix for proof). For Bernoulli random variables,
\(X\) and \(Y\) are unfaithful if and only if \(\rho = 0\) and
\(\gamma \neq 0\).

The following model is a simple example where \(X\) and \(Y\) are
unfaithful. Let \(\epsilon_Z, \epsilon_Y\) be fair coins. Then, in the
following model, \(\rho=0\) and \(\gamma=1/2\):

\[Z = \epsilon_Z\] \[X = Z\] \[
Y =
\begin{cases}
\lnot Z, & \text{if } \epsilon_Y = 1 \\
X, & \text{if } \epsilon_Y = 0
\end{cases}
\]

\begin{longtable}[]{@{}llll@{}}
\caption{Observational distribution of unfaithful model}\tabularnewline
\toprule
\(P(x,y)\) & y=0 & y=1 & \(P(x)\)\tabularnewline
\midrule
\endfirsthead
\toprule
\(P(x,y)\) & y=0 & y=1 & \(P(x)\)\tabularnewline
\midrule
\endhead
x=0 & 1/4 & 1/4 & 1/2\tabularnewline
x=1 & 1/4 & 1/4 & 1/2\tabularnewline
\(P(y)\) & 1/2 & 1/2 &\tabularnewline
\bottomrule
\end{longtable}

\begin{longtable}[]{@{}lll@{}}
\caption{Interventional distributions of unfaithful
model}\tabularnewline
\toprule
\(P(y \mid \hat{x})\) & y=0 & y=1\tabularnewline
\midrule
\endfirsthead
\toprule
\(P(y \mid \hat{x})\) & y=0 & y=1\tabularnewline
\midrule
\endhead
x=0 & 3/4 & 1/4\tabularnewline
x=1 & 1/4 & 3/4\tabularnewline
\bottomrule
\end{longtable}

Almost all models are faithful in a formal sense -- models that do not
respect the faithfulness condition have Lebeguse measure zero in
probability spaces where model parameters have continuous support and
are independently distributed \autocite{spirtes2000}. However, this does
not mean that such models can be dismissed out of hand; they are
vanishingly unlikely to occur by chance, but can be deliberately
engineered.

\subsubsection{Friedman's thermostat and the traitorous
lieutenant}\label{friedmans-thermostat-and-the-traitorous-lieutenant}

Consider ``Friedman's Thermostat''; a correctly functioning thermostat
would keep the indoor temperature constant, regardless of the external
temperature, by adjusting the furnace settings.\footnote{Friedman
  introduced the thermostat analogy in the context of a central bank
  controlling money supply \autocite{friedman2003}. Its use as a general
  analogy for correlation and causation has been popularized by Rowe
  \autocite{rowe2010}.} Observation would show external temperature and
furnace settings to be anticorrelated with each other and internal
temperature to be uncorrelated with both. This does not correspond to
the true causal effect that external temperature and furnace settings
have on internal temperature.

The sharp-eyed reader will note that the Friedman's thermostat example
is not a recursive (acyclic) causal model. An example of unfaithfulness
with a recursive causal model can be seen in the following ``Traitorous
Lieutenant'' problem. Consider the problem of a general trying to send a
one-bit message. The general has two lieutenants available to act as
messengers, however, one of them is a traitor and will leak whatever
information they have to the enemy. The general observes the following
protocol: to send a \(1\), the general either gives the first lieutenant
a \(1\) and the other a \(0\), or the first a \(0\) and the second a
\(1\), with equal probability. To send a \(0\), the general either gives
both lieutenants a \(0\), or both lieutenants a \(1\), with equal
probability. The recipient of the message XORs both lieutenants' bits to
recover the original message.

\begin{figure}[htbp]
\centering
\includegraphics{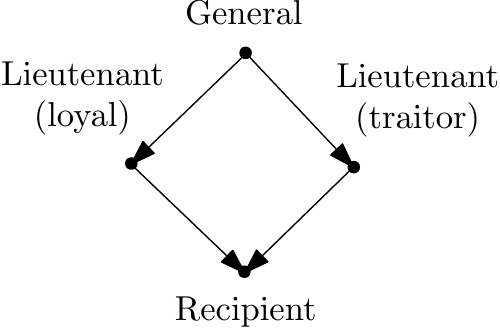}
\caption{Diagram of the traitorous lieutenant problem}
\end{figure}

In this scenario, the traitor will see a \(0\) and \(1\) with equal
probability, regardless of the actual message. This is unfaithfulness; a
lieutenant changing their bit has a causal effect on the final message,
but observing a single lieutenant's bit provides no information.

\subsection{Genuine causation and confounding
bias}\label{genuine-causation-and-confounding-bias}

``Genuine causation'' is suggested for referring to models where
\(\rho\) and \(\gamma\) have the same sign. However, due to confounding
bias, the strength of the true casual effect may be different than what
a naive interpretation of correlation would suggest.

The causal definition of \emph{no confounding} is provided by Pearl
\autocite{pearl1998}.

\begin{description}
\tightlist
\item[No confounding]
Let \(M\) be a causal model. \(X\) and \(Y\) are not confounded in \(M\)
if and only if \(P(y \mid \hat{x}) = P(y \mid x)\).
\end{description}

By the definition of the causation coefficient, no confounding implies
\(\rho=\gamma\).

Genuine causation with negative confounding bias corresponds to
\(\gamma > \rho\), and can be thought of as a weaker version of the type
of confounding effect that produces unfaithfulness or inverse causation.
In such cases, the true causal effect will be stronger than correlation
suggests. Let \(\epsilon_Z\) be a fair coin and \(\epsilon_Y\) be
Bernoulli distributed with \(p=1/4\). Then, the following model exhibits
genuine causation with negative confounding bias, with \(\rho=1/2\) and
\(\gamma=3/4\):

\[Z = \epsilon_Z\] \[X = Z\] \[
Y =
\begin{cases}
\lnot Z, & \text{if } \epsilon_Y=1 \\
X, & \text{if } \epsilon_Y=0
\end{cases}
\]

\begin{longtable}[]{@{}llll@{}}
\caption{Observational distribution of negative confounding bias
model}\tabularnewline
\toprule
\(P(x,y)\) & y=0 & y=1 & \(P(x)\)\tabularnewline
\midrule
\endfirsthead
\toprule
\(P(x,y)\) & y=0 & y=1 & \(P(x)\)\tabularnewline
\midrule
\endhead
x=0 & 3/8 & 1/8 & 1/2\tabularnewline
x=1 & 1/8 & 3/8 & 1/2\tabularnewline
\(P(y)\) & 1/2 & 1/2 &\tabularnewline
\bottomrule
\end{longtable}

\begin{longtable}[]{@{}lll@{}}
\caption{Interventional distributions of negative confounding bias
model}\tabularnewline
\toprule
\(P(y \mid \hat{x})\) & y=0 & y=1\tabularnewline
\midrule
\endfirsthead
\toprule
\(P(y \mid \hat{x})\) & y=0 & y=1\tabularnewline
\midrule
\endhead
x=0 & 7/8 & 1/8\tabularnewline
x=1 & 1/8 & 7/8\tabularnewline
\bottomrule
\end{longtable}

Genuine causation with positive confounding bias corresponds to
\(\gamma < \rho\); in such cases, the true causal effect will be weaker
than correlation suggests. Let \(\epsilon_X, \epsilon_Y, \epsilon_Z\) be
fair coins. In the following model, \(\rho \approx 0.745\), the natural
causation coefficient, \(\gamma \approx 0.447\), and the maximum entropy
causation coefficient, \(\gamma_H = 0.5\):

\[Z = \epsilon_Z\] \[X = Z \land \epsilon_X\] \[
Y =
\begin{cases}
Z, & \text{if } \epsilon_Y=1 \\
X, & \text{if } \epsilon_Y=0
\end{cases}
\]

\begin{longtable}[]{@{}llll@{}}
\caption{Observational distribution of positive confounding bias
model}\tabularnewline
\toprule
\(P(x,y)\) & y=0 & y=1 & \(P(x)\)\tabularnewline
\midrule
\endfirsthead
\toprule
\(P(x,y)\) & y=0 & y=1 & \(P(x)\)\tabularnewline
\midrule
\endhead
x=0 & 5/8 & 1/8 & 3/4\tabularnewline
x=1 & 0 & 1/4 & 1/4\tabularnewline
\(P(y)\) & 5/8 & 3/8 &\tabularnewline
\bottomrule
\end{longtable}

\begin{longtable}[]{@{}lll@{}}
\caption{Interventional distributions of positive confounding bias
model}\tabularnewline
\toprule
\(P(y \mid \hat{x})\) & y=0 & y=1\tabularnewline
\midrule
\endfirsthead
\toprule
\(P(y \mid \hat{x})\) & y=0 & y=1\tabularnewline
\midrule
\endhead
x=0 & 3/4 & 1/4\tabularnewline
x=1 & 1/4 & 3/4\tabularnewline
\bottomrule
\end{longtable}

\section{Typical relationship between correlation and
causation}\label{typical-relationship-between-correlation-and-causation}

Common intuition suggests that correlation is closely related to
causation. However, the models in the previous section act as a
constructive proof that the sign of the correlation coefficient provides
no guarantees about the true causal effect. Some insight on this
apparent discrepancy can be found by considering the following set of
linear probabilistic causal models, parameterized by
\(\sigma_{\epsilon_X}^2, \sigma_{\epsilon_Y}^2, \sigma_{\epsilon_Z}^2, \alpha_Z, \beta_X, \beta_Z\):

\[Z = \epsilon_Z\] \[X = \alpha_Z Z + \epsilon_X\]
\[Y = \beta_X X + \beta_Z Z + \epsilon_Y\]

Since these models are linear and covariance is bilinear, the population
correlation coefficient can be calculated analytically, regardless of
the underlying distribution of the error terms:

\[
\rho_{X, Y} = \frac{\beta_X \sigma_{\epsilon_X}^2 + (\alpha_Z^2 \beta_X + \alpha_Z \beta_Z) \sigma_{\epsilon_Z}^2}{\sqrt{(\sigma_{\epsilon_X}^2 + \alpha_Z^2 \sigma_{\epsilon_Z}^2) (\beta_X^2 \sigma_{\epsilon_X}^2 + \sigma_{\epsilon_Y}^2 + (\alpha_Z \beta_X + \beta_Z)^2 \sigma_{\epsilon_Z}^2)}}
\]

The natural causation coefficient can also be calculated directly from
the definitions of the causation coefficient and causal effect:

\[\gamma_{X \to Y} = \frac{\beta_X \sigma_{\epsilon_X}^2 + \alpha_Z^2 \beta_X \sigma_{\epsilon_Z}^2}{\sqrt{(\sigma_{\epsilon_X}^2 + \alpha_Z^2  \sigma_{\epsilon_X}^2) (\beta_X^2 \sigma_{\epsilon_X}^2 + \sigma_{\epsilon_Y}^2 + (\alpha_Z^2 \beta_X^2 + \beta_Z^2) \sigma_{\epsilon_Z}^2)}}\]

\begin{figure}[htbp]
\centering
\includegraphics{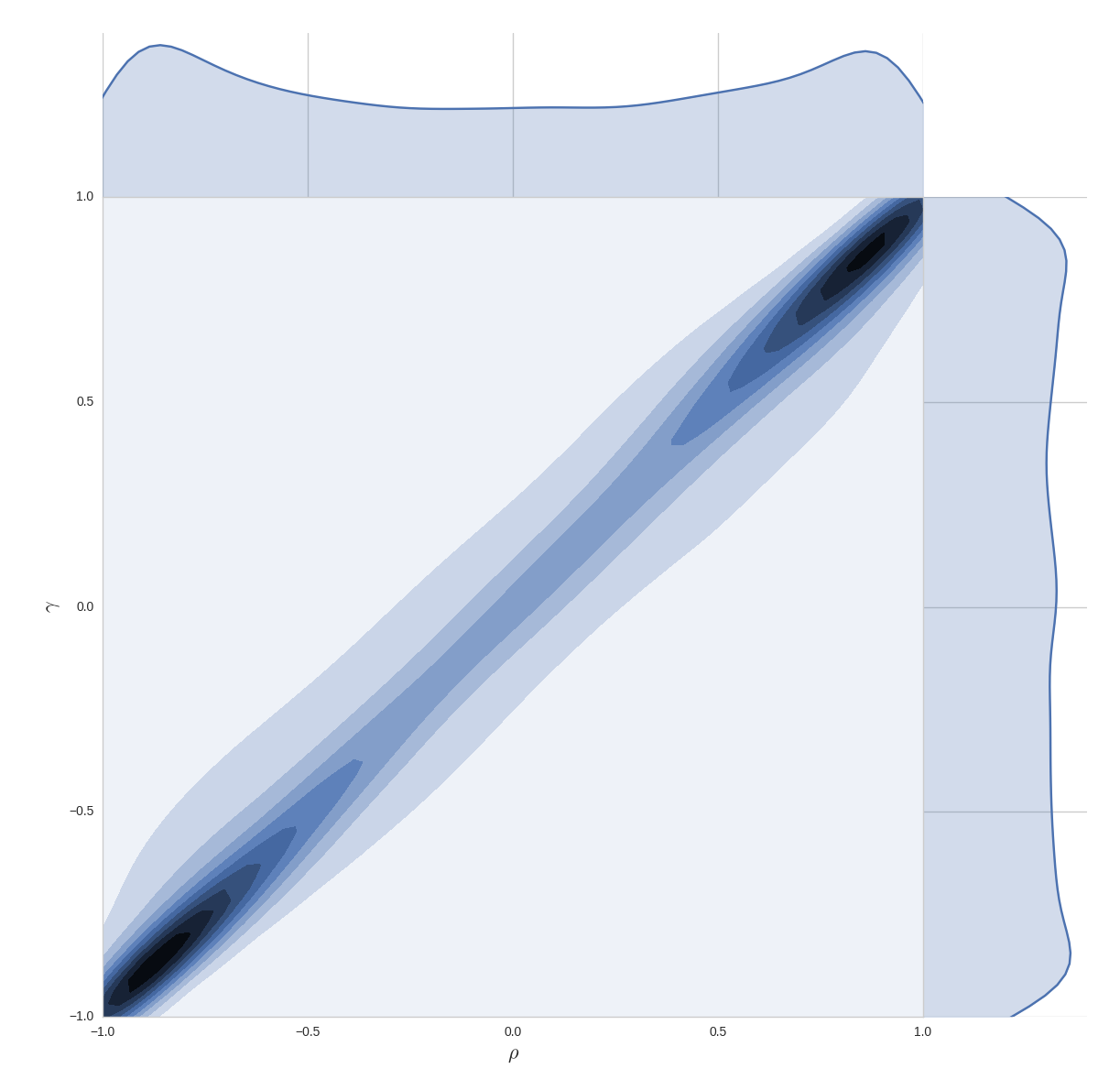}
\caption{Causation vs correlation coefficient (kernel density estimation
with \(10^6\) samples). Darker shading indicates higher density of
models. The curves at the top and right of the graph are the marginal
densities of \(\rho\) and \(\gamma\). \label{kde}}
\end{figure}

The typical relationship between correlation and causation can be
analyzed by constructing a probability distribution for the parameters
of the linear model. \(\alpha_Z, \beta_X, \beta_Z\) have support over
the entire real line;
\(\sigma_{\epsilon_X}^2, \sigma_{\epsilon_Y}^2, \sigma_{\epsilon_Z}^2\)
have support over \((0, \infty)\). Given mean \(0\) and variance \(1\),
the maximum entropy distributions are \(N(0,1)\) and \(\text{exp}(1)\),
respectively. Assuming jointly independent distributions over the
parameters, it is straightforward to randomly sample models and compute
their correlation and causation coefficients. Plotting \(\gamma\)
against \(\rho\) yields a graph where each point represents a single
linear probabilistic causal model. The (smoothed) result of plotting
such a graph is in Figure \ref{kde}.

In the graph of \(\gamma\) vs \(\rho\), the upper left and lower right
quadrants contain inverse causation models and the other two quadrants
contain genuine causation models. Except for \((0,0)\), which
corresponds to an invariant and independent model, the horizontal line,
\(\rho=0\), contains common causation models and the vertical line,
\(\gamma=0\), contains unfaithful models.

With maximum entropy distributions over the parameters, the probability
of a random linear model exhibiting inverse causation \(\approx 0.122\),
genuine causation with negative bias \(\approx 0.364\), and genuine
causation with positive bias \(\approx 0.514\). This matches closely
with common intuition. Typically, a strong positive correlation
indicates a strong positive causal effect -- this can be seen in the
upper right quadrant, with a high density of models. Inverse causation
is possible, although much less likely, and unfaithful models have
measure \(0\), which accounts for why they are often considered
counterintuitive. However, this is an analysis of population, not sample
coefficients and the measure of nearly unfaithful models is
nonzero.\footnote{Formally, the measure of \(\lambda\)-strong-unfaithful
  distributions converges to 1 exponentially in the number of nodes
  \autocite{uhler2013}.} In practice, this means that unfaithfulness
cannot be dismissed as irrelevant. Although the population correlation
will be zero in such models, the sample correlation will often be
indistinguishable from zero, despite the possibility of a nontrivial
causal effect.

The choice of a maximum entropy distribution in this analysis is based
on the principle of maximum entropy, which states that the appropriate
prior distribution, given the absence of any other information, is the
maximum entropy distribution \autocite{jaynes2003}. However, the choice
of linear models and the particular parameterization remain somewhat
subjective. The statement that inverse causation only occurs in
\(\approx 12\%\) of models should be seen as qualitatively consistent
with the intuition that such situations are rare, but not quantitatively
significant.

\section{Estimating the causal
coefficient}\label{estimating-the-causal-coefficient}

Randomization of an independent variable effectively cuts all incoming
edges to that node in a causal diagram, removing potential confounding
effects. Reporting a correlation coefficient, in the context of a
randomized controlled trial, can be viewed as reporting an estimate of
the causation coefficient, with the distribution of interventions,
\(\hat{P}(x)\), equal to the distribution of interventions that were
performed in the experiment.

When randomization is not available, it may still be possible calculate
a sample causal coefficient by estimating \(P(y \mid do(x))\). Presented
here is a simple example, using data from a study on the treatment of
kidney stones \autocite{charig1986}. More advanced techniques for
identifying \(P(y \mid do(x))\) are given in \autocite{shpitser2008}.

The subgroups (\(Z\)) in Table \ref{kidney} refer to kidney stone size.
Group 1 is small kidney stones; group 2 is large kidney stones. This
study can be modeled with binary treatment (\(X\)) and response (\(Y\))
variables, with the decision to perform percutaneous nephrolithotomy
(PCNL) as \(0\) and surgery as \(1\).

\begin{longtable}[]{@{}llll@{}}
\caption{Success rate of treatment; successful/total (probability)
\label{kidney}}\tabularnewline
\toprule
& Group 1 & Group 2 & \emph{Overall}\tabularnewline
\midrule
\endfirsthead
\toprule
& Group 1 & Group 2 & \emph{Overall}\tabularnewline
\midrule
\endhead
Open surgery & 81/87 (0.93) & 192/263 (0.73) & \emph{273/350
(0.78)}\tabularnewline
PCNL & 234/270 (0.87) & 55/80 (0.69) & \emph{289/350
(0.83)}\tabularnewline
\emph{Overall} & \emph{315/357 (0.88)} & \emph{247/343 (0.72)} &
\textbf{562/700 (0.80)}\tabularnewline
\bottomrule
\end{longtable}

\begin{figure}[htbp]
\centering
\includegraphics{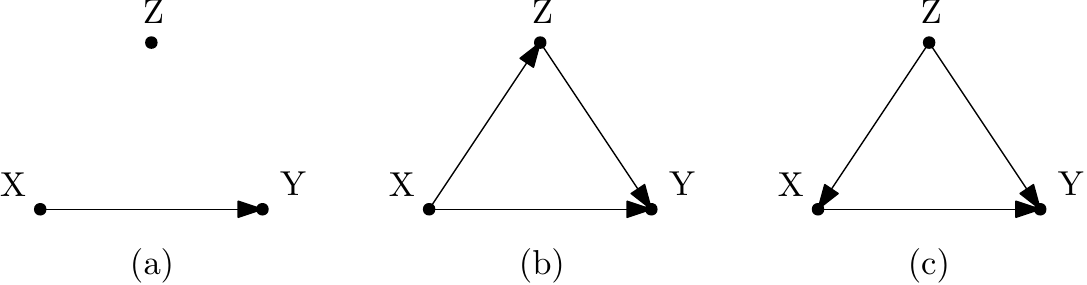}
\caption{Some of the possible causal diagrams for modeling kidney stone
treatment \label{kidney-diagrams}}
\end{figure}

The naive model is that there is no confounding (Figure
\ref{kidney-diagrams}a). In such a case, the population natural
causation coefficient equals the population correlation coefficient and
therefore the sample correlation coefficient \(r\) is equal to the
sample causation coefficient \(g\).\footnote{Statisticians normally
  denote an estimate of a parameter, \(\theta\), with a hat symbol,
  \(\hat{\theta}\). This convention is not followed here, since the hat
  symbol has been used to indicate causal effect.} Given the data,
\(r=g=-0.057\). The cohorts are equal in size, so this is also an
estimate of the maximum entropy causation coefficient.

Hypothetically, if the subgroups were postoperative infection and the
treatment affected the likelihood of postoperative infection, which, in
turn, affected recovery (Figure \ref{kidney-diagrams}b), the natural
causation coefficient would still equal the correlation coefficient --
adjusting for subgroups would still be incorrect. This is an immediate
consequence of the do-calculus \autocite{shpitser2008}.

However, the correct set of causal assumptions is that kidney stone size
affects treatment and recovery (Figure \ref{kidney-diagrams}c) --
doctors took kidney stone size into account when making the decision
whether or not to send a patient to surgery. Correctly estimating the
causation coefficient in this model can be done with an adjustment for
direct causes, \(P(y \mid do(x)) = \sum_z P(y \mid x, z) P(z)\). With
respect to the correct causal diagram, estimating the causation
coefficient yields \(g=0.068\). This a case of inverse causation and the
best treatment option for patients is the opposite of what a naive
interpretation of correlation would suggest.

The reversal effect seen here is well known as Simpson's paradox, but
requires causal knowledge to resolve correctly \autocite{pearl2014}.
Adjusting for the wrong variables will produce incorrect estimates of
causal effect.

\section{Conclusions}\label{conclusions}

There are many different ways in which positive correlation can be
misleading with respect to causation. Population distributions may
exhibit common causation (\(\rho > 0, \gamma = 0\)) or inverse causation
(\(\rho > 0, \gamma < 0\)). Sampling error introduces the possibility of
coincidental correlation (\(r > 0, \rho = 0\)). Unfaithfulness
(\(\rho = 0, \gamma \neq 0\)) implies that the absence of correlation
cannot guarantee the absence of causation. And even if there is no
confounding (\(\rho = \gamma\)), human error introduces the possibility
of reverse causation (\(\gamma_{X \to Y} \neq \gamma_{Y \to X}\)).

Despite the warning that, ``Correlation is not causation'', the two are
easy to conflate because of the high likelihood that a random model will
have \(\rho \approx \gamma\). However, there remains a nontrivial
possibility of encountering other correlation/causation relationships
such as inverse causation, a problem that no amount of additional data
sampling will mitigate. There is simply no substitute for accurate
causal assumptions.

By emphasizing the population/sample and statistical/causal distinctions
and explicitly naming the different ways in which correlation can relate
to causation, it is hoped that these effects will become easier to
recognize in practice.

\section{Appendix}\label{appendix}

\textbf{Theorem.} \quad For Bernoulli \(X\), \(Y\),
\(\gamma_{X \to Y} = 0\) if and only if \(Y\) is invariant to \(X\).

\textbf{Proof.} Consider the definition of average treatment effect,
\(\text{ATE}(X \to Y) = P(y=1 \mid do(x=1)) - P(y=1 \mid do(x=0))\).
Average treatment effect is zero if and only if
\(P(y=1 \mid do(x=1)) = P(y=1 \mid do(x=0))\). Since the support of a
Bernoulli random variable is \(\{0, 1\}\), this is equivalent to \(Y\)
invariant to \(X\). Since \(\gamma\) has the same sign as the average
treatment effect, \(\gamma_{X \to Y}=0\) if and only if \(Y\) is
invariant to \(X\). \(\square\)

\textbf{Theorem.} \quad For Bernoulli \(X\), \(Y\):
\(\gamma_{X \to Y} = \text{ATE}(X \to Y) (Var[\hat{X}] / Var_{\hat{X}}[Y])^{1/2}\)

\textbf{Proof.} \quad Consider the numerator of \(\gamma\). For
Bernoulli random variables:

\begin{align*}
P(y \mid do(x=1)) \hat{P}(x=1) - \hat{P}(x=1) (& P(y=1 \mid do(x=1)) \hat{P}(x=1) \\
&+ P(y=1 \mid do(x=0)) \hat{P}(x=0) )
\end{align*}\begin{align*}
=\hat{P}(x=1) ( P(y=1 \mid do(x=1)) &- \hat{P}(x=1) P(y=1 \mid do(x=1)) \\
&- \hat{P}(x=0) P(y=1 \mid do(x=0)))
\end{align*}\begin{align*}
= \hat{P}(x=1) (& P(y=1 \mid do(x=1)) - \hat{P}(x=1) P(y=1 \mid do(x=1)) \\
&- (1 - \hat{P}(x=1)) P(y=1 \mid do(x=0)))
\end{align*}\begin{align*}
= \hat{P}(x=1) (& P(y=1 \mid do(x=1)) - P(y=1 \mid do(x=0)) \\
                &- \hat{P}(x=1) (P(y=1 \mid do(x=1)) - P(y=1 \mid do(x=0)) ))
\end{align*}\begin{align*}
= &(P(y=1 \mid do(x=1)) - P(y=1 \mid do(x=0)) ) \hat{P}(x=1) (1 - \hat{P}(x=1)) \\
= &\text{ATE}(X \to Y) Var[\hat{X}]
\end{align*}

Therefore,
\(\gamma_{X \to Y} = \text{ATE}(X \to Y)(Var[\hat{X}] / Var_{\hat{X}}[Y])^{1/2}\).
\(\square\)

\textbf{Theorem.} \quad In all recursive probabilistic causal models, at
least one of \(\gamma_{X \to Y}, \gamma_{Y \to X}\) is zero.

\textbf{Proof.} \quad Assume without loss of generality that
\(\gamma_{x \to y}\) is nonzero. This implies that \(Y\) is not
invariant to \(X\). Since \(P(y \mid \hat{x})\) is a nonconstant
function of \(x\), \(X\) must be an ancestor of \(Y\) in the associated
causal diagram. Consider the submodel \(M_y\) that results from
\(do(Y=y)\). Since, \(X\) is an ancestor of \(Y\) in the original model,
\(X\) and \(Y\) must be d-separated in \(M_y\). Therefore, \(X\) is
invariant to \(Y\) and \(\gamma_{Y \to X}\) is zero. \(\square\)

\textbf{Theorem.} \quad If \(X\) and \(Y\) are unfaithful in causal
model \(M\), then the observational distribution \(P\) and causal
diagram \(G\) associated with \(M\) violate the faithfulness condition.

\textbf{Proof.} \quad Assume without loss of generality that \(Y\) is
not invariant to \(X\). Therefore, \(X\) is an ancestor of \(Y\) in the
associated causal diagram and \(X\) and \(Y\) are d-connected
\autocite{pearl2009}. However, \(X\) and \(Y\) are independent, an
independence relation not entailed by the Markov condition. Therefore
the observational distribution \(P\) is not faithful to \(G\).
\(\square\)

\section{Acknowledgements}\label{acknowledgements}

Thanks to James Reggia, Brendan Good, Donald Gregorich and Richard Bruns
for their comments on drafts of this paper.

\printbibliography[title=References]

\end{document}